\documentclass[12pt]{iopart}
\usepackage{flushend}
\usepackage{graphicx}
\usepackage{dcolumn}
\usepackage{appendix}
 \usepackage{upgreek}
\usepackage{float}
\usepackage{subfigure}
\usepackage[
bookmarks=true,
colorlinks,
linkcolor=blue,
urlcolor=blue,
citecolor=blue,
plainpages=false,
pdfpagelabels,
final,
breaklinks=true
]{hyperref}

\begin{document}

\title{Dynamic laser ablation loading of a linear Paul trap}

\author{Lin Li$^{1,2}$,Zi Li$^{1,2}$,Xia Hua$^{1}$ and Xin Tong$^{1,3,*}$}

\address{$^1$ Innovation Academy for Precision Measurement Science and Technology Chinese Academy of Sciences, Wuhan 430071, China}
\address{$^2$ University of Chinese Academy of Sciences, Beijing 100049, China}
\address{$^3$ Wuhan Institute of Quantum Technology, Wuhan, 430206, China}
\address{$^*$ Author to whom any correspondence should be addressed.}
\ead{tongxin@wipm.ac.cn}

\vspace{10pt}

\begin{abstract}
We present a detailed method for accumulating Ca$^{+}$ ions controllably in a linear Paul trap. The ions are generated by pulsed laser ablation and dynamically loaded into the ion trap by switching the trapping potential on and off. The loaded ions are precooled by buffer gas and then laser-cooled to form Coulomb crystals for verifying quantity. The number of ions is controlled by manipulating the trapping potential of the ion trap, partial pressure of buffer gas and turn-on time of the entrance end cap voltage. With single-pulse laser ablation, the number of trapped ions ranges from tens to ten thousand. The kinetic energy of loaded ions can be selected via the optimal turn-on time of the entrance end cap. Using multiple-pulse laser ablation, the number is further increased and reaches about $4 \times 10^{4}$. The dynamic loading method has wide application for accumulating low-yielding ions via laser ablation in the ion trap.

\end{abstract}

\begin{indented}
\item[]\textbf{ Keywords:} laser ablation, linear Paul trap, dynamic ion loading, ions Coulomb crystal
\end{indented}

\section{Introduction}

Trapped atomic ions are widely used in experiments and applications, such as optical frequency standards \cite{1,2}, precision measurements of fundamental physics \cite{3,4}, and quantum computer \cite{5,6}. A conventional method for loading ions into an ion trap consists of the production and ionization of atoms. A neutral flux of atoms is commonly produced from a resistively heated oven, and atoms are ionized in the trap via electron impact \cite{7,8} or photoionization \cite{9,10}. The heated oven is typically operated for tens of seconds and at temperatures over several hundred Kelvin, resulting in a significant heating background and limited application in a cryogenic environment. The electron impact may also deposit charges on insulators, resulting in changes of electric potentials. Furthermore, the neutral atomic flux is difficult to be produced for the elements that aggressively react with background gases or have extremely high melting points \cite{11}.\par 
Laser ablation can offer an alternative method for loading ions into an ion trap \cite{12}. When a short-pulse laser with sufficient fluence impacts on a target sample, different atomic species including neutral atoms, singly and multiply charged ions can be produced \cite{13,14,15}. Here, the laser fluence is the laser energy divided by spot area. With the low fluence of the laser pulse, neutral atoms are predominantly produced during the ablation process \cite{osada2023compact} and can be further photoionized in an ion trap \cite{16,17,18,48,christensen2020high}. While the fluence of the laser pulse is high enough \cite{19}, a large number of charged ions and electrons are produced during the ablation process \cite{20,21,22,23,24,25}. The high-flux electrons can interfere with the radio frequency (rf) electric field in the ion trap and the trapping potential drops for a short amount of time to allow the ions entering the trap \cite{25}. When the potential recovers, the ions in the trapping region are confined. At present, laser ablation with the high fluence of the laser pulse is mainly employed for loading slow and singly charged ions \cite{22,23,24,25}. However, it may also cause the ion trap unstable due to the deposition of the charges on the trap electrodes \cite{23}. \par
In the paper, we present a detailed method for accumulating ions controllably in a linear Paul trap and preventing the contamination of trap electrodes by the ablated charged particles. The ions are produced via pulsed laser ablation. The trapping potential is actively switched off once the ablation laser fires and then turned back when the ablated ions reach the trapping region \cite{46,47}. Compared to photoionization, this dynamic laser ablation loading method cannot selectively load isotopes, subsequent isotope selection via secular motion excitation \cite{Vedel1991, March1997} or selectively laser heating \cite{Toyoda2001, Kitaoka2012} is required after the loading of ions. However, this dynamic laser ablation loading method allows trapping not only slow and singly charged ions but also fast and multi-charged ions. We demonstrate the ion loading efficiency with Ca$^{+}$ ions under various rf voltages, end cap voltages, buffer gas partial pressures and turn-on times of the entrance end cap. Besides, ion loading and accumulating with multiple-pulse laser ablation are investigated under different energies of the ablation laser.  \par 

\section{ Experiment}
In the experiment, a linear Paul trap for the accumulation of Ca$^{+}$ ions is housed in a vacuum chamber with a background pressure lower than 1 $\times$ ${10^{-10}}$ mbar. As shown in Fig. 1(a), the trap consists of four stainless-steel rods of 80 mm length and 8 mm diameter. The radial trapping potential is created by a 2 MHz rf voltage ($V_{\mathrm{rf}}=50-600$ V$_{\mathrm{pp}}$) applied to all four rods. Confinement in the axial direction is provided by a dc voltage ($U_{\rm{end}}=10-200$ V) on the stainless-steel end cap electrodes separated by 20 mm and insulted from the rf rods by polyimide tapes. Then the axial trap potential depth can be obtained by $D_{\rm{axial}}=\kappa QU_{\rm{end}} $, where $\kappa=0.106$ is the geometric parameters of the ion trap in the axial direction, $Q$ is the charge of the trapped ions. Both end caps have 6 mm coaxial holes to the trap axis for ion loading and optical access. The ion trap is fixed onto a bow-shaped stainless-steel bracket by two stainless-steel square holders with 6 mm coaxial holes. The two stainless-steel square holders are insulated from the steel rods through the aluminum oxide ceramic rings. \par

A calcium target (purity $\textgreater$ 99$\%$, thickness $\sim$ 0.4 mm) is chosen for demonstration of the laser ablation and the loading into a linear Paul trap. The target is attached on a stainless-steel plate and located next to a 2 mm diameter coaxial hole to the trap axis as shown in Fig. 1(a). The distance from the calcium target to the entrance end cap is 55 mm. A diode-pumped pulsed laser at wavelength 1064 nm, with the energy of $\sim$ 110 $\upmu$J/pulse and 8 ns duration, has a Gaussian spatial profile and is focused to a 1/e$^2$ diameter of 80(10) $\upmu$m on a calcium target. The ablation laser energy is monitored by an energy meter (Coherent, FieldMaxII) and the laser is operated at a stable condition but with a large energy of about 20 mJ. Then the laser passes through a neutral density filter (Thorlabs, NDUV513B) for energy attenuation, and the energy is controlled by varying the size of the diaphragm. Therefore, the final energy fluctuation is small, with a fluctuation of about 10 $\upmu$J at the laser energy of 70 $\upmu$J. The laser impacts the calcium target at the incident angle of approximately 30{\textdegree}, and then the Ca$^+$ ions are generated. The ions pass through the grounded stainless-steel square holder with a hole in the center and are further guided by trap electrodes to minimize the charge depositing on the electrodes. \par

\begin{figure}[!htpb]
\centering
\includegraphics[width=14cm]{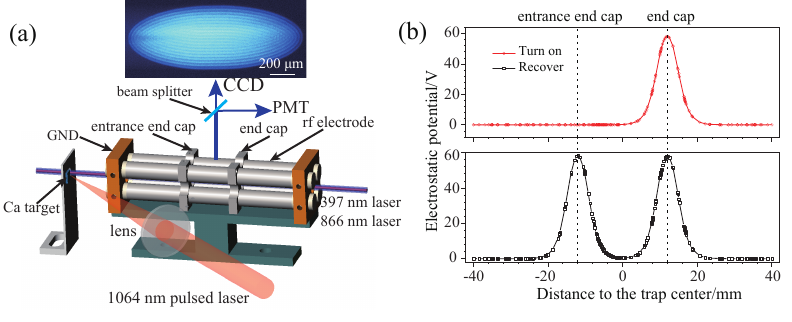}
\caption{\label{fig1} Experimental apparatus and operation for calcium ion loading. (a) A linear Paul trap is used for ion trapping. The ablation laser is used for the generation of ions; the 397 nm and 866 nm lasers are used for Doppler laser cooling, and their wavelengths are monitored and locked with a wavemeter (WSU-10). The fluorescence of ions is guided through the beam splitter and detected by a CCD and a PMT. (b) Axial electrostatic potential distribution is calculated as a function of distance to the trap center; turn on (in red): the voltage of the entrance end cap is pulsed down from 70 V to 0 V; recover (in black): the voltage of the entrance end cap is turned back from 0 V to 70 V. Turn-on time of entrance end cap: time duration of holding 0 V on the entrance end cap. The dotted line indicates the center position of the end caps.}
\end{figure} \par

To realize dynamic loading ions into the ion trap, before the ablation laser fires, both end caps are kept at an axial trapping voltage of $\sim$ 70 V. When the ablation laser fires, the voltage on the entrance end cap is switched off from 70 V to 0 V within 1 $\upmu$s and kept at 0 V for a while as shown in Fig. 1(b). When the ablated ions reach the trap center, the voltage on the entrance end cap rises to 70 V within 1 $\upmu$s to realize the axial confinement.  \par 

To decelerate the trapped ions, helium buffer gas with 99.9999$\%$ (volume percent) purity is introduced into the vacuum chamber via a manual leak valve (Kurt Lesker, VZLVM267) for collisionally cooling of the trapped ions to room temperature. After the buffer gas cooling, the 397 nm cooling laser and 866 nm repumping laser are employed to laser-cool the trapped ions below 6.5 mK \cite{hornekaer2002formation}. Both lasers are set to the power of 1 mW and the beam waist of 1 mm. The 397 nm laser is red detuned by 20 MHz from the resonant frequency of the dipole transition $4 s$ $^2S_{1/2} \leftrightarrow 4p$ $^2P_{1/2}$. The 866 nm laser is resonant on the transition of $4p$ $^2P_{1/2} \leftrightarrow 3 d$ $^2D_{3/2}$ \cite{27}. The fluorescence of the cold ions is split equally and guided into a Charge-coupled Device (CCD, Andor iKon-M 934) and a photomultiplier tube (PMT, Thorlabs PMT1002). The trapped and laser-cooled ions form an ordered structure, so-called ion Coulomb crystal (ICC) as shown in Fig. 1(a) \cite{26}. The typical image of ICC obtained by the CCD is shown in Fig. 1(a). The number density of the ICC is calculated under the zero-temperature approximation \cite{37}. The shape of the ICC is approximated as an ellipsoid, and the volume of the ICC is obtained by fitting the ellipse to the contour of the ion fluorescence image. Therefore, the absolute number of the ICC is obtained by the product of the number density and volume of the ICC. The fluorescence of ions at constant partial pressure (typically $10^{-6}$ mbar) of He gas is detected by the PMT and the photon counts represent the relative ion number in the different ICCs.\par

\section{Results and discussion}
\subsection{Ion loading with single-pulse ablation}
In the experiment, the ion loading efficiency with single-pulse ablation is investigated under various rf voltages, end cap voltages, buffer gas partial pressures and turn-on times of the entrance end cap. \par
\begin{figure}[!htpb]
\centering
\includegraphics[width=14cm]{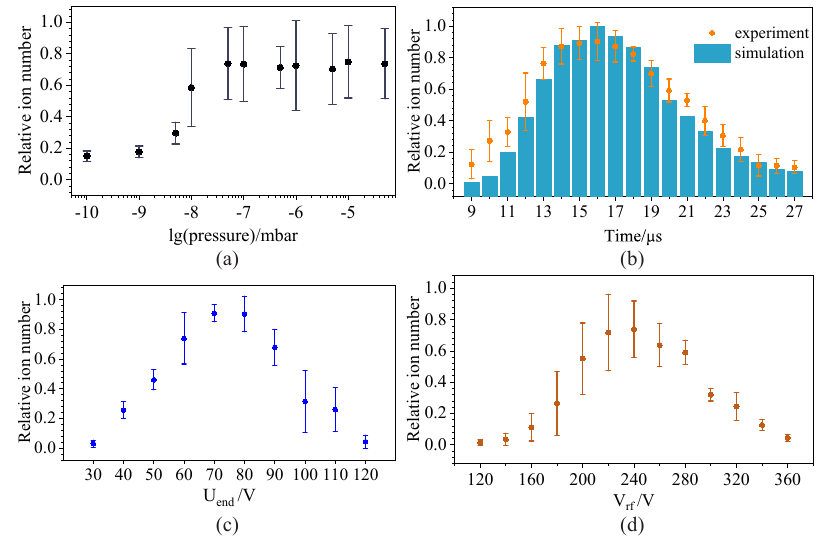}
\caption{\label{fig:4}Relative ion number of trapped ions are obtained (a) as a function of the partial pressure of the He gas, where $V_{\rm{rf,p-p}}=240$ V, $U_{\rm{end}}=70$ V, the turn-on time of the entrance end cap is set to 16 $\upmu$s. After the ions are loaded at different partial pressures of He gas, the partial pressure of He gas is set back to 10$^{-6}$ mbar for PMT counts measurements. The pressure values in the abscissa are scaled logarithmically, (b) as a function of the turn-on time of the entrance end cap, where $V_{\rm{rf,p-p}}=240$ V, $U_{\rm{end}}=70$ V, the partial pressure of the He gas is set to 10$^{-6}$ mbar, the scatterplot of solid orange dots indicates the results of the experiment and the blue vertical bar plot represents simulation results, (c) as a function of the entrance end cap voltage, where $V_{\rm{rf,p-p}}=240$ V, the partial pressure of He gas is set to 10$^{-6}$ mbar and the turn-on time of the entrance end cap is set to 16 $\upmu$s. After the ions are loaded at different end cap voltages, the end cap voltage is set back to 70 V for PMT counts measurements. (d) as a function of the rf voltage, where $U_{\rm{end}}=70$ V, the partial pressure of He gas is set to 10$^{-6}$ mbar and the turn-on time of the entrance end cap is set to 16 $\upmu$s. After the ions are loaded at different rf voltages, the rf voltage is set back to 240 V for PMT counts measurements. Each point in Fig (a-d) represents an average of five independent experiments, while the error bars in the figures indicate the standard deviation of the mean.}
\end{figure}\par
The ablated ions loaded into the trap are unstable due to their high velocities \cite{28}. The He gas is introduced as the buffer gas to collisionally cool the ions. To investigate the influence of the buffer gas cooling on loading ions, the partial pressure of He gas is varied from $10^{-10}$ mbar to $5 \times 10^{-5}$ mbar. At the buffer gas partial pressure of 5 $\times$ $10^{-5}$ mbar, the ion-gas collision frequency is about 0.001/$\upmu$s \cite{31}. The ablated ions with an average velocity of $ \sim $ 10 mm/$\upmu$s \cite{29,30} take about 2 $\upmu$s traveling from the entrance end cap to the other end cap. It takes about 1.9 s to sufficiently cool the ion cloud with buffer gas at partial pressure of 10$^{-6}$ mbar \cite{delahaye2019analytical}. Therefore, the He gas can only cool the trapped ions with sufficient trapping time, and the ions with only a single passing through the ion trap can not be effectively cooled \cite{32,33}. As shown in Fig. 2(a), the number of loaded ions is increased as the partial pressure of He gas is changed from $10^{-10}$ mbar to $5 \times 10^{-8}$ mbar due to the increasing collisional rate. When the partial pressure of He gas is in the range of $5 \times10^{-8}$ mbar to $5 \times 10^{-5}$ mbar, the trapped ions are sufficiently collisional cooled and the number of loaded ions is saturated.\par

The ablated ions have a large velocity distribution \cite{34}, the turn-on time of the entrance end cap is required to match the time that ions arrive at the trap center. To investigate the influence of turn-on time on the efficiency of ion loading, the turn-on time is varied from 9 $\upmu$s to 27 $\upmu$s as shown in Fig. 2(b). When the turn-on time of the entrance end cap is below 9 $\upmu$s or above 27 $\upmu$s, no fluorescence of trapped ions can be detected. When the turn-on time is chosen to be 16 $\upmu$s, the maximum number of ions is loaded. It confirms that the most probable velocity distribution of ions that can be loaded is at the turn-on time of 16 $\upmu$s. The ablated ions follow the shifted Maxwell-Boltzmann-Coulomb distribution \cite{34}, and a 3D Monte Carlo simulation of the dynamic loading is performed. As a result, the velocity distribution of ablated ions is obtained when the center-of-mass and Coulomb velocities are set to be 0.91 mm/$\upmu$s and 0.49 mm/$\upmu$s, respectively. The simulated ion number loaded with different turn-on times is in line with the experimental results shown in Fig. 2(b). It means that the kinetic energy of loaded ions can be selected by manipulating the turn-on time of the entrance end cap by our dynamic loading method. \par

At the turn-on time of 16 $\upmu$s, the slow ions with the kinetic energy of 2.41 eV travel just from the ablation target to the entrance end cap. The fast ions with the kinetic energy of 7.36 eV pass through the entrance end cap to the other end cap, and then they are repelled back to the entrance end cap. To investigate the dependence of ion loading efficiency on axial trapping potentials, the end cap voltage is varied from 30 V to 120 V which corresponds to the axial potential depth of 3.18 eV to 12.72 eV. As shown in Fig. 2(c), when the end cap voltage is varied from 30 V to 70 V, the number of loaded ions increases due to the increasing axial potential depth. The efficiency of ion loading is at the maximum when the end cap voltage is set to 70 V, because the potential depth is greater than the kinetic energy of the fast ions that can be loaded. When the end cap voltage is over 80 V, the number of loaded ions decreases because the trapping condition becomes unstable with the increased end cap voltage \cite{25,36}. \par

The trap parameter $q$ is proportional to the rf voltage \cite{35} and determines the stability of the trapped ions, which in turn influences the loading efficiency of the ions. To investigate the dependence of ion loading efficiency on the stability parameter $q$, the applied rf voltage is varied from 120 V to 360 V, corresponding to the stability parameter $q$ from 0.2 to 0.9, while the end cap voltage is set to 70 V for the optimized loading condition as described above. As shown in Fig. 2(d), when the rf voltage is below 120 V or above 360 V, no fluorescence of trapped ions can be detected. The efficiency of ion loading is the highest when the rf voltage is set to 240 V, the corresponding $q$ of 0.603 is consistent with the stability parameter of the most stable trapped ions \cite{25,36}. \par

As the above studies concluded, with the optimal trap parameter of $V_{\rm{rf,p-p}}=240$ V and $U_{\rm{end}}=70$ V, the 16 $\upmu$s turn-on time of entrance end cap and the He gas partial pressure of $10^{-6}$ mbar, Ca$^{+}$ ions can be readily loaded into the ion trap by single-pulse laser with the energy of 110 $\upmu$J. The number of ions can be determined to $1.26(1) \times 10^{4}$ by the image of an ICC generated by laser cooling as shown in Fig. 1(a), and the uncertainty of the ion number comes from the estimation of the crystal volume. The numbers of loaded ions at turn-on time between 18 and 25 $\upmu$s range from hundreds to tens of thousands in Fig. 2(b). Five independent experiments were performed at each turn-on time, and about 5$\%$ uncertainty is the statistical error coming from the ion numbers of all the loading experiments.\par

Compared to directly loading ions into an ion trap via laser ablation without switching trapping potential \cite{22,23,24,25}, the number of loaded ions by dynamic laser ablation loading can be manipulated not only via rf voltage and end cap voltage but also via the turn-on time of the entrance end cap and the partial pressure of buffer gas. Furthermore, the ions with specific velocity distribution can be selectively loaded using the different turn-on times of the entrance end cap.  \par

\subsection{Ion loading with multiple-pulse laser ablation }
Accumulation of ions in the ion trap can be realized with several times of switching the voltage of the entrance end cap on and off during the multiple-pulse laser ablation. The relative loading efficiencies with the laser pulse energy range from 70 $\upmu$J to 210 $\upmu$J are shown in Fig. 3(a). Each trial contains 8 successive ablation laser pulses with 3-second intervals. Due to the frequency locking of the laser and the stabilization of the laser power, the fluctuations in the fluorescence emitted by the ions are small enough to ignore the error bars. As the energy of the ablation laser increases, the relative loading efficiencies for the first pulse ablation in each trial increase gradually. \par 

\begin{figure}[htp]
\centering
\includegraphics[width=14cm]{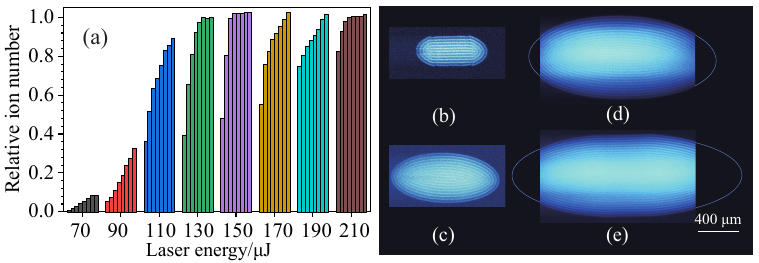}
\caption{\label{fig4} Ions are loaded via 8-pulse ablation. (a) Relative PMT counts of trapped ions in the He gas of $10^{-6}$ mbar via 8 pulses' ablation with different pulse energy. Four ICCs are obtained via 8 pulses' ablation with pulse energy of (b) 70 $\upmu$J, (c) 90 $\upmu$J, (d) 110 $\upmu$J and (e) 130 $\upmu$J, where $V_{\rm{rf,p-p}}=240$ V, $U_{\rm{end}}=70$ V, the 16 $\upmu$s turn-on time of entrance end cap and the He gas partial pressure of $10^{-6}$ mbar. The images of ICCs in (d) and (e) are too large to be fitted within the CCD sensor, The solid blue lines represent the fitted boundaries of the ICCs.}
\end{figure} \par

When the pulse energy of the ablation laser is below 70(10) $\upmu$J, no ion is loaded into the ion trap due to the low fluence of the ablation laser. The ablation threshold for successfully loading ions is then obtained to be 1.4(3) J/cm$^2$ with the ablation laser spot size of 80(10) $\upmu$m, which is consistent with the ablation threshold of metal targets \cite{39}. \par
At the low energy range of 70 and 90 $\upmu$J of the ablation laser, the numbers of Ca$^{+}$ ions loaded and accumulated in the ion trap increase with almost equal quantities for each ablation laser pulse. After 8 laser pulses, the final ICCs are formed as shown in Fig. 3(b) and 3(c) for 70 and 90 $\upmu$J pulse energy, respectively.\par

When the pulse energy of the ablation laser is 110 $\upmu$J, the majority of ions are accumulated via the first few ablation pulses. Although the consequent loadings do not reach saturation, they contribute much less newly accumulated ions. This is due to stronger Coulomb repulsion created by the trapped high-density ions preventing further accumulation. The final ICC is formed as shown in Fig. 3(d). The number of ions in the ICC is estimated to be 3.89(20) $\times$ 10$^4$. \par

When the pulse energy of the ablation laser is over 130 $\upmu$J, ion loading and accumulation are more efficient, however, the number of trapped ions is saturated as shown in Fig. 3(a). A typical final ICC containing $4.68(13) \times 10^{4}$ Ca$^{+}$ ions is formed as shown in Fig. 3(e). One can improve the accuracy of the loaded-ion numbers with an even lower-energy ablation laser and more ablation pulses. In contrast, the efficiency of ion loading can be increased by a high-energy ablation laser and fewer pulses while sacrificing the accuracy.\par

Apart from dynamic multiple-pulse laser ablation loading, the ions can also be accumulated in the ion trap via direct multiple-pulse laser ablation loading without switching the trapping potential. 20 ions were accumulated in the ion trap via direct 120-pulse laser ablation with fluence of 1.4 J/cm$^{2}$ \cite{24}. While using dynamic multiple-pulse laser ablation loading, the number of loaded ions can range from several to tens of thousands within several pulses. The maximum number of loaded ions is three orders of magnitude greater than with direct loading because the fluence of the ablation laser can have a wider adjustable range. In addition, the number of loaded ions is more controllable because more parameters (the rf voltage and end cap voltage of the ion trap, the partial pressure of He gas, the energy of ablation laser and the turn-on time of the entrance end cap) can be manipulated. \par

\section{Conclusion}
In this paper, we present a detailed method for accumulating ions controllably in a linear Paul trap via laser ablation and switching the voltage of the entrance end cap on and off. The relative loading efficiency against the partial pressure of He gas, rf voltage and end cap voltage of the ion trap, and turn-on time of the entrance end cap has been investigated. A large Coulomb crystal consisting of $1.26(1) \times 10^{4}$ Ca$^{+}$ ions is obtained via single-pulse laser ablation. By multiple switching voltage of the entrance end cap on and off during multiple-pulse laser ablation, the loaded ions can be accumulated controllably in the ion trap and the final number of the ions can reach $4.68(13) \times 10^{4}$.  \par

Although the direct loading using multiple-pulse laser ablation has successfully accumulated a small number ($\textless100$) of ions in the ion trap without actively changing the trapping potential \cite{24}, our multiple-pulse dynamic laser ablation loading method has two unique advantages. First, the maximum number of loaded ions is three orders of magnitude greater than with direct loading. Second, the number of loaded ions is more controllable.\par

We have demonstrated that over 10 thousand ions can be accumulated controllably via single-pulse and multiple-pulse dynamic laser ablation loading. This method can be applied in the fields of precision measurement and spectroscopy requiring large-number ($\textgreater1000$) ion systems to enhance high signal-to-noise ratios. For instance, large-number ions can be used in the development of microwave ions clocks such as $^{113}$Cd$^{+}$ \cite{40,49}, $^{171}$Yb$^{+}$ \cite{41}, $^{199}$Hg$^{+}$ \cite{42} and $^{9}$Be$^{+}$ \cite{43}. An ion ICC consists of large-number ions that can be used to be the coolant for sympathetic cooling atom ions \cite{44} and molecular ions \cite{45}. Furthermore, the kinetic energy of loaded ions can be selected via the optimal turn-on time of the entrance end cap, and the low kinetic energy of ions can be loaded into a trap with low trapping potential.\par

\section*{Data availability statement}
All data that support the findings of this study are included
within the article (and any supplementary files).

\ack{}
This work was supported by the National Key R$\&$D Program of China (Grant No. 2021YFA1402103) and the National Natural Science Foundation of China (Grant No. 11804372).

\section*{Declaration of competing interest}
The authors declare that they have no known competing
financial interests or personal relationships that could have
appeared to influence the work reported in this paper.

\section*{Ethical approval}
There are no ethical issues involved in this study.


\nocite{*}

\section*{References }

\begin{thebibliography}{10}

\bibitem{4}
S~Alighanbari, GS~Giri, Florin~Lucian Constantin, VI~Korobov, and Stephan Schiller.
\newblock Precise test of quantum electrodynamics and determination of fundamental constants with {HD}$^{+}$ ions.
\newblock {\em Nature}, 581(7807):152--158, 2020.

\bibitem{39}
Salvatore Amoruso, Riccardo Bruzzese, Nicola Spinelli, and R~Velotta.
\newblock Characterization of laser-ablation plasmas.
\newblock {\em J. Phys. B: At. Mol. Opt. Phys.}, 32(14):R131, 1999.

\bibitem{7}
Z~Andelkovic, R~Cazan, W~N{\"o}rtersh{\"a}user, S~Bharadia, DM~Segal, RC~Thompson, R~J{\"o}hren, J~Vollbrecht, V~Hannen, and M~Vogel.
\newblock Laser cooling of externally produced {Mg} ions in a penning trap for sympathetic cooling of highly charged ions.
\newblock {\em Phys. Rev. A}, 87(3):033423, 2013.

\bibitem{12}
PB~Antohi, D~Schuster, GM~Akselrod, J~Labaziewicz, Y~Ge, Z~Lin, WS~Bakr, and IL~Chuang.
\newblock Cryogenic ion trapping systems with surface-electrode traps.
\newblock {\em Rev. Sci. Instrum.}, 80(1):013103, 2009.

\bibitem{42}
D.~J. Berkeland, J.~D. Miller, J.~C. Bergquist, W.~M. Itano, and D.~J. Wineland.
\newblock Laser-cooled mercury ion frequency standard.
\newblock {\em Phys. Rev. Lett.}, 80:2089--2092, Mar 1998.

\bibitem{43}
JJ~Bollinger and J~Heinzen.
\newblock A 303-mhz frequency standard based on trapped.
\newblock {\em IEEE Trans. Instrum. Meas.}, 40(2), 1991.

\bibitem{34}
Petr~V Borisyuk, Sergey~P Derevyashkin, Ksenia~Y Khabarova, Nikolay~N Kolachevsky, Yury~Y Lebedinsky, Sergey~S Poteshin, Alexey~A Sysoev, Evgeny~V Tkalya, Dmitry~O Tregubov, Viktor~I Troyan, et~al.
\newblock Loading of mass spectrometry ion trap with th ions by laser ablation for nuclear frequency standard application.
\newblock {\em Eur. J. Mass Spectrom.}, 23(4):146--151, 2017.

\bibitem{2}
S.~M. Brewer, J.-S. Chen, A.~M. Hankin, E.~R. Clements, C.~W. Chou, D.~J. Wineland, D.~B. Hume, and D.~R. Leibrandt.
\newblock $^{27}${Al}$^{+}$ quantum-logic clock with a systematic uncertainty below ${10}^{\ensuremath{-}18}$.
\newblock {\em Phys. Rev. Lett.}, 123:033201, Jul 2019.

\bibitem{30}
C.~J. Campbell.
\newblock {\em Trapping, laser cooling and spetroscopy of thorium {IV}}.
\newblock Thesis, Georgia Institute of Technology, 2011.

\bibitem{47}
CJ~Campbell, AG~Radnaev, and A~Kuzmich.
\newblock Wigner crystals of $^{229}${Th} for optical excitation of the nuclear isomer.
\newblock {\em Phys. Rev. Lett.}, 106(22):223001, 2011.

\bibitem{46}
CJ~Campbell, AV~Steele, LR~Churchill, MV~DePalatis, DE~Naylor, DN~Matsukevich, A~Kuzmich, and MS~Chapman.
\newblock Multiply charged thorium crystals for nuclear laser spectroscopy.
\newblock {\em Phys. Rev. Lett.}, 102(23):233004, 2009.

\bibitem{christensen2020high}
Justin~E Christensen.
\newblock {\em High-Fidelity Operation of a Radioactive Trapped-Ion Qubit, $^{133}${Ba}$^+$}.
\newblock Thesis, University of California, 2020.

\bibitem{28}
LS~Cutler, RP~Giffard, and MD~McGuire.
\newblock Thermalization of $^{199}${Hg} ion macromotion by a light background gas in an rf quadrupole trap.
\newblock {\em Appl. Phys. B}, 36(3):137--142, 1985.

\bibitem{5}
Shantanu Debnath, Norbert~M Linke, Caroline Figgatt, Kevin~A Landsman, Kevin Wright, and Christopher Monroe.
\newblock Demonstration of a small programmable quantum computer with atomic qubits.
\newblock {\em Nature}, 536(7614):63--66, 2016.

\bibitem{delahaye2019analytical}
Pierre Delahaye.
\newblock Analytical model of an ion cloud cooled by collisions in a paul trap.
\newblock {\em The European Physical Journal A}, 55(5):83, 2019.

\bibitem{33}
SP~Derevyashkin, PV~Borisyuk, K~Yu Khabarova, NN~Kolachevsky, SA~Strelkin, EV~Tkalya, DO~Tregubov, IV~Tronin, and VP~Yakovlev.
\newblock Cumulative loading of the ion trap by laser ablation of thorium target in buffer gas.
\newblock {\em Laser Phys. Lett.}, 18(1):015501, 2020.

\bibitem{19}
S~Di~Mihai, A~Marcu, and N~Puscas.
\newblock Pulsed laser ablation of solids: basics, theory and applications, 2014.

\bibitem{31}
Li~Ding, Michael Sudakov, and Sumio Kumashiro.
\newblock A simulation study of the digital ion trap mass spectrometer.
\newblock {\em Int. J. Mass Spectrom.}, 221(2):117--138, 2002.

\bibitem{44}
JZ~Han, HR~Qin, NC~Xin, YM~Yu, VA~Dzuba, JW~Zhang, and LJ~Wang.
\newblock Toward a high-performance transportable microwave frequency standard based on sympathetically cooled $^{113}${Cd}$^{+}$ ions.
\newblock {\em Appl. Phys. Lett.}, 118(10):101103, 2021.

\bibitem{25}
Yoshinori Hashimoto, Leo Matsuoka, Hiroyuki Osaki, Yu~Fukushima, and Shuichi Hasegawa.
\newblock Trapping laser ablated {Ca}$^{+}$ ions in linear {Paul} trap.
\newblock {\em Jpn. J. Appl. Phys.}, 45(9R):7108, 2006.

\bibitem{37}
L~Hornek{\ae}r, N~Kj{\ae}rgaard, AM~Thommesen, and M~Drewsen.
\newblock Structural properties of two-component coulomb crystals in linear paul traps.
\newblock {\em Phys. Rev. Lett.}, 86(10):1994, 2001.

\bibitem{hornekaer2002formation}
Liv Hornek{\ae}r and M~Drewsen.
\newblock Formation process of large ion coulomb crystals in linear paul traps.
\newblock {\em Physical Review A}, 66(1):013412, 2002.

\bibitem{1}
Yao Huang, Baolin Zhang, Mengyan Zeng, Yanmei Hao, Zixiao Ma, Huaqing Zhang, Hua Guan, Zheng Chen, Miao Wang, and Kelin Gao.
\newblock Liquid-nitrogen-cooled {Ca}$^{+}$ optical clock with systematic uncertainty of 3$\times$ 10$^{-18}$.
\newblock {\em Phys. Rev. Appl.}, 17(3):034041, 2022.

\bibitem{36}
R~Iffl{\"a}nder and G~Werth.
\newblock Optical detection of ions confined in a rf quadrupole trap.
\newblock {\em Metrologia}, 13(3):167, 1977.

\bibitem{40}
B.~M. Jelenkovi\ifmmode~\acute{c}\else \'{c}\fi{}, S.~Chung, J.~D. Prestage, and L.~Maleki.
\newblock High-resolution microwave-optical double-resonance spectroscopy of hyperfine splitting of trapped $^{113}\mathrm{Cd}^{+}$ ions.
\newblock {\em Phys. Rev. A}, 74:022505, Aug 2006.

\bibitem{32}
A~Kellerbauer, T~Kim, RB~Moore, and P~Varfalvy.
\newblock Buffer gas cooling of ion beams.
\newblock {\em Nucll. Instrum. Meth. A}, 469(2):276--285, 2001.

\bibitem{Kitaoka2012}
Masami Kitaoka and Shuichi Hasegawa.
\newblock Isotope-selective manipulation of ca+ using laser heating and cooling in a linear paul trap.
\newblock {\em Journal of Physics B: Atomic, Molecular and Optical Physics}, 45, 2012.

\bibitem{45}
JCJ Koelemeij, B~Roth, A~Wicht, I~Ernsting, and S~Schiller.
\newblock Vibrational spectroscopy of {HD}$^{+}$ with 2-ppb accuracy.
\newblock {\em Phys. Rev. Lett.}, 98(17):173002, 2007.

\bibitem{3}
IV~Kortunov, S~Alighanbari, MG~Hansen, GS~Giri, VI~Korobov, and S~Schiller.
\newblock Proton--electron mass ratio by high-resolution optical spectroscopy of ion ensembles in the resolved-carrier regime.
\newblock {\em Nat. Phys.}, 17(5):569--573, 2021.

\bibitem{22}
Victor~HS Kwong.
\newblock Production and storage of low-energy highly charged ions by laser ablation and an ion trap.
\newblock {\em Phys. Rev. A}, 39(9):4451, 1989.

\bibitem{23}
David~R Leibrandt, Robert~J Clark, Jaroslaw Labaziewicz, Paul Antohi, Waseem Bakr, Kenneth~R Brown, and Isaac~L Chuang.
\newblock Laser ablation loading of a surface-electrode ion trap.
\newblock {\em Phys. Rev. A}, 76(5):055403, 2007.

\bibitem{27}
H.~X. Li, Y.~Zhang, S.~G. He, and X.~Tong.
\newblock Determination of the geometric parameters $\kappa_{r}$ and $\kappa_{z}$ of a linear paul trap.
\newblock {\em Chinese J. Phys.}, 60:61--67, 2019.

\bibitem{10}
Min Li, Yong Zhang, Qian-Yu Zhang, Wen-Li Bai, Sheng-Guo He, Wen-Cui Peng, and Xin Tong.
\newblock An efficient method for producing $^{9}${Be}$^{+}$ ions using a 2+1 resonance-enhanced multiphoton ionization process.
\newblock {\em J. Phys. B: At. Mol. Opt. Phys.}, 55(3):035002, 2022.

\bibitem{6}
Yao Lu, Shuaining Zhang, Kuan Zhang, Wentao Chen, Yangchao Shen, Jialiang Zhang, Jing-Ning Zhang, and Kihwan Kim.
\newblock Global entangling gates on arbitrary ion qubits.
\newblock {\em Nature}, 572(7769):363--367, 2019.

\bibitem{9}
DM~Lucas, A~Ramos, JP~Home, MJ~McDonnell, S~Nakayama, J-P Stacey, SC~Webster, DN~Stacey, and AM~Steane.
\newblock Isotope-selective photoionization for calcium ion trapping.
\newblock {\em Phys. Rev. A}, 69(1):012711, 2004.

\bibitem{March1997}
Raymond~E March.
\newblock An introduction to quadrupole ion trap mass spectrometry.
\newblock {\em Journal of Mass Spectrometry}, 32:351--369, 1997.

\bibitem{49}
SN~Miao, HR~Qin, NC~Xin, JZ~Han, YT~Chen, JW~Zhang, and LJ~Wang.
\newblock Sympathetic cooling of a large $^{113}${Cd}$^+$ ion crystal with $^{40}${Ca}$^+$ in a linear {Paul} trap.
\newblock {\em Chinese J. Phys.}, 83:242--252, 2023.

\bibitem{41}
S~Mulholland, HA~Klein, GP~Barwood, S~Donnellan, D~Gentle, G~Huang, G~Walsh, PEG Baird, and P~Gill.
\newblock Laser-cooled ytterbium-ion microwave frequency standard.
\newblock {\em Appl. Phys. B}, 125(11):1--12, 2019.

\bibitem{20}
S~Olmschenk and P~Becker.
\newblock Laser ablation production of {Ba}, {Ca}, {Dy}, {Er}, {La}, {Lu}, and {Yb} ions.
\newblock {\em Appl. Phys. B}, 123(4):1--6, 2017.

\bibitem{osada2023compact}
A~Osada, R~Tamaki, W~Lin, I~Nakamura, and A~Noguchi.
\newblock Compact strontium atom source using fiber-based pulsed laser ablation.
\newblock {\em Applied Physics Letters}, 122(18), 2023.

\bibitem{48}
Alto Osada and Atsushi Noguchi.
\newblock Deterministic loading of a single strontium ion into a surface electrode trap using pulsed laser ablation.
\newblock {\em Journal of Physics Communications}, 6(1):015007, 2022.

\bibitem{35}
Wolfgang Paul.
\newblock Electromagnetic traps for charged and neutral particles.
\newblock {\em Rev. Mod. Phys.}, 62:531--540, Jul 1990.

\bibitem{8}
Vladimir~L Ryjkov, XianZhen Zhao, and Hans~A Schuessler.
\newblock Sympathetic cooling of fullerene ions by laser-cooled {Mg}$^{+}$ ions in a linear rf trap.
\newblock {\em Phys. Rev. A}, 74(2):023401, 2006.

\bibitem{21}
Muhammed Sameed, Daniel Maxwell, and Niels Madsen.
\newblock Ion generation and loading of a penning trap using pulsed laser ablation.
\newblock {\em New J. Phys.}, 22(1):013009, 2020.

\bibitem{16}
H~Shao, M~Wang, M~Zeng, H~Guan, and K~Gao.
\newblock Laser ablation and two-step photo-ionization for the generation of $^{40}${Ca}$^{+}$.
\newblock {\em J. Phys. Commun.}, 2(9):095019, 2018.

\bibitem{15}
SN~Srivastava, BK~Sinha, and KP~Rohr.
\newblock Ions and ion-energy spectra of a collisional laser plasma produced from multi-species targets of aluminium and titanium.
\newblock {\em J. Phys. B: At. Mol. Opt. Phys.}, 39(14):3073, 2006.

\bibitem{29}
B~Thestrup, B~Toftmann, J{\o}rgen Schou, B~Doggett, and JG~Lunney.
\newblock Ion dynamics in laser ablation plumes from selected metals at 355 nm.
\newblock {\em Appl. Surf. Sci}, 197:175--180, 2002.

\bibitem{26}
Richard~C Thompson.
\newblock Ion coulomb crystals.
\newblock {\em Contemp. Phys.}, 56(1):63--79, 2015.

\bibitem{13}
L~Torrisi, F~Caridi, D~Margarone, and L~Giuffrida.
\newblock Nickel plasma produced by 532-nm and 1064-nm pulsed laser ablation.
\newblock {\em Plasma Phys. Rep.}, 34(7):547--554, 2008.

\bibitem{Toyoda2001}
Kenji Toyoda, Hiroshi Kataoka, Yasushi Kai, Akihiko Miura, Masayoshi Watanabe, and Shinji Urabe.
\newblock Separation of laser-cooled 42ca+ and 44ca+ in a linear paul trap.
\newblock {\em Applied Physics B}, 72:327--330, 2001.

\bibitem{Vedel1991}
Fernande Vedel, Michel Vedel, and Raymond~E March.
\newblock A sensitive method for the detection of stored ions by resonant ejection using a wide-band signal.
\newblock {\em International Journal of Mass Spectrometry and Ion Processes}, 108, 1991.

\bibitem{17}
Geert Vrijsen, Yuhi Aikyo, Robert~F Spivey, I~Volkan Inlek, and Jungsang Kim.
\newblock Efficient isotope-selective pulsed laser ablation loading of $^{174}${Yb}$^{+}$ ions in a surface electrode trap.
\newblock {\em Opt. Express}, 27(23):33907--33914, 2019.

\bibitem{18}
Brendan~M White, Pei~Jiang Low, Yvette De~Sereville, Matthew~L Day, Noah Greenberg, Richard Rademacher, and Crystal Senko.
\newblock Isotope-selective laser ablation ion-trap loading of $^{137}${Ba}$^{+}$ using a {BaCl}$_{2}$ target.
\newblock {\em Phys. Rev. A}, 105(3):033102, 2022.

\bibitem{24}
Qiming Wu, Melina Filzinger, Yue Shi, Zhihui Wang, and Jiehang Zhang.
\newblock Adaptively controlled fast production of defect-free beryllium ion crystals using pulsed laser ablation.
\newblock {\em Rev. Sci. Instrum.}, 92(6):063201, 2021.

\bibitem{14}
JJ~Zhang, HY~Zhao, GC~Wang, LT~Sun, XZ~Zhang, GP~Li, and HW~Zhao.
\newblock Ion charge state and energy distributions of laser produced plasma from pure metals and their alloy.
\newblock {\em Rev. Sci. Instrum.}, 90(12):123306, 2019.

\bibitem{11}
K~Zimmermann, MV~Okhapkin, OA~Herrera-Sancho, and E~Peik.
\newblock Laser ablation loading of a radiofrequency ion trap.
\newblock {\em Appl. Phys. B}, 107(4):883--889, 2012.

\end{thebibliography}
\end{document}